# Multimodal Nonlinear Microscope based on a Compact Fiber-format Laser Source

*Francesco Crisafi[1]\*, Vikas Kumar[1]\*, Antonio Perri[1], Marco Marangoni[1], Giulio Cerullo[1], Dario Polli[1,2]†*

1. IFN-CNR, Dipartimento di Fisica, Politecnico di Milano, Piazza Leonardo da Vinci 32, 20133, Milano, Italy
2. Center for Nano Science and Technology @Polimi, Istituto Italiano di Tecnologia, via Giovanni Pascoli 70/3, 20133 Milano, Italy

\* *Authors contributed equally to work*   † *Corresponding author: dario.polli@polimi.it*

**Abstract**

We present a multimodal non-linear optical (NLO) laser-scanning microscope, based on a compact fiber-format excitation laser and integrating coherent anti-Stokes Raman scattering (CARS), stimulated Raman scattering (SRS) and two-photon-excitation fluorescence (TPEF) on a single platform. We demonstrate its capabilities in simultaneously acquiring CARS and SRS images of a blend of 6-μm poly(methyl methacrylate) beads and 3-μm polystyrene beads. We then apply it to visualize cell walls and chloroplast of an unprocessed fresh leaf of *Elodea* aquatic plant via SRS and TPEF modalities, respectively. The presented NLO microscope, developed in house using off-the-shelf components, offers full accessibility to the optical path and ensures its easy re-configurability and flexibility.

## 1 Introduction

Optical microscopy is an extremely powerful investigation tool in life sciences, thanks to its ability of visualizing morphological details in cells and tissues on the sub-micrometer spatial scale [1]. It provides a much higher spatial resolution compared to magnetic resonance imaging, and, at the same time, it does not require the sample to be fixed, as in electron microscopy. Nonlinear optical (NLO) microscopy techniques, in particular, offer additional advantages, such as inherent 3D-sectioning capability and greater penetration depth, due to the use of infrared wavelengths. The most widespread NLO technique is two-photon-excitation fluorescence (TPEF) [2-4]. It provides very high sensitivity, but in most cases it requires the addition of markers to the sample to be studied, either exogenous (staining dyes or semiconductor quantum dots) or endogenous (fluorescent proteins). Other NLO microscopy techniques have the advantage of being label-free, allowing the use of pristine samples. Some of these techniques, such as second-harmonic generation (SHG) [5-7], sum-frequency generation (SFG) [8, 9] and third-harmonic generation (THG) [4, 10] microscopy, do not provide chemical contrast, i.e. they are not able to selectively differentiate between specific components of a cell or a tissue. SHG observes contrast in non-centrosymmetric structures like collagen fibers in tissue, while THG is mainly sensitive to local changes of nonlinear refractive index within the sample occurring at interfaces, such as lipid cell membranes or lipid droplets. Chemical selectivity is provided by coherent Raman scattering (CRS) techniques [11], which employ the vibrational spectrum of a molecule to provide an endogenous signature that can be used for its identification. CRS exploits the third-order nonlinear optical response of the sample to set up and detect a vibrational coherence within the ensemble of molecules inside the laser focus. When the difference between pump and Stokes frequencies matches a characteristic vibrational frequency, then all the molecules in the focal volume are resonantly excited and vibrate in phase; this vibrational coherence enhances the Raman response by many orders of magnitude with respect to the incoherent spontaneous Raman process.

CRS microscopy has two popular forms: coherent anti-Stokes Raman scattering (CARS) [12, 13] and stimulated Raman scattering (SRS) [14-17]. In both CARS and SRS, two synchronized narrowband pulses, the pump (at frequency $\omega_p$) and the Stokes (at frequency $\omega_s$), are focused on a sample and their frequency difference is tuned to a Raman-active vibrational mode $\Omega$ of the targeted molecule, i.e. $\Omega = \omega_p - \omega_s$. When it happens, in the case of CARS, a strong anti-Stokes signal at frequency $\omega_{as} = 2\omega_p - \omega_s$ is generated, which is utilized to probe the molecules under study. Since CARS signal is at higher frequency with respect to input pump-Stokes frequencies, it can be easily spectrally separated and is also immune to one-photon fluorescence signals which fall at lower frequencies. This makes CARS signal naturally free from any kind of linear background. On the other hand, the CARS process has the serious drawback that it suffers from nonlinear non-resonant background (NRB). CARS signal is given by $I_{CARS} \propto |\chi^{(3)}|^2 = |\chi_R^{(3)}(\Omega) + \chi_{NR}^{(3)}|^2$, where $\chi_R^{(3)}(\Omega)$ is the complex resonant response of the targeted vibration

and $\chi_{NR}^{(3)}$ is the real non-resonant response of the background, which does not deliver any chemically specific information. In SRS, on the other hand, the coherent interaction with the sample induces stimulated emission from a virtual state of the sample to the investigated vibrational state, resulting in a Stokes-field amplification (stimulated Raman gain, SRG) and in a simultaneous pump-field attenuation (stimulated Raman loss, SRL). SRS is inherently free from NRB, however it is technically demanding as it requires the detection of a weak signal (the SRG/SRL) on top of a large and fluctuating linear background. This typically requires sophisticated modulation-transfer techniques.

Multimodal NLO microscopy can capitalize the potential of different NLO modalities by combining two or more of them in a single imaging platform. Hence, it can provide richer microscopic information by imaging different kinds of molecules or structures in a sample. Multimodal CARS/TPEF/SHG microscopes have already proven for the capability of biological imaging and identification of cancerous tissues in brain [18], lung [19] and kidney [20]. Cheng *et al.* have utilized the combination of TPEF/SFG/CARS modalities to investigate the central nervous system in diseased and healthy states [21], in the study of the progression of arterial diseases [22] and in imaging and quantitative analysis of atherosclerosis, a major cause of cardiovascular diseases [6]. The combination of THG/SHG/TPEF microscopies has been used e.g. to visualize the microstructure of human cornea [4].

The main stumbling block which prevents widespread adoption of multimodal NLO microscopy techniques in the biological and medical communities is the complication and cost of the experimental apparatus, both for the excitation laser and the microscope. Regarding the excitation source, the most critical technique is CRS. A compromise must be found between high peak laser power to enhance the nonlinear signal, which points towards shorter pulses, and the frequency resolution to preserve molecular selectivity, which points towards narrow laser bandwidth. In the condensed phase, Raman transitions exhibit linewidths of the order of tens of $cm^{-1}$, so that the optimal pulse duration lies in the 1-3 picosecond range for pump/Stokes pulses. Moreover, their frequency difference should be tunable, to access a large vibrational bandwidth and the laser repetition rate should be high ($\approx$100 MHz) to reach the shot-noise limit in detection and avoid multi-photon absorption sample damage caused by the high peak power of the pulses. An output power of $\approx$100 mW per branch is also typically required to compensate for losses in the optical chain of the microscope. The complexity of the excitation laser is one reason why, after its early demonstration in 1982 [23], the development of CRS microscopy has stopped for nearly two decades. Following its revival, initial CRS implementations were based on two electronically synchronized picosecond Ti:sapphire oscillators [24-26], while the current "gold standard" is represented by an optical parametric oscillator synchronously pumped by a picosecond Nd:YVO$_4$ oscillator [27-29]. Such systems are complex, expensive, and they all critically require a synchronization between two independent laser sources, which must be maintained over time. Drastically simplified excitation architectures, with lower cost and smaller footprint, are thus greatly in demand, as they would lower the technological entrance barriers to CRS microscopy. In addition, the multimodal NLO microscope platform cannot easily rely on commercially available solutions. The detection of the different nonlinear signals, in fact, requires full accessibility to the optical paths, in order to place different components and detectors required to implement the various techniques. This is difficult in commercial microscope systems that, being designed for the end user, are typically not accessible and modifiable.

In this paper, we describe a multimodal NLO laser-scanning microscope with a highly simplified architecture, based on a compact fiber-based laser, which enables the TPEF, CARS and SRS modalities. The fiber laser generates two synchronized beams of picosecond pulses, the pump at 780 nm and the Stokes tunable between 950 and 1050 nm, with sufficient spectral coverage to implement CARS and SRS microscopy in the C-H stretching region. A single excitation beam can also be used for other NLO microscopy modalities, such as TPEF, SHG, THG and SFG. The excitation source is coupled to a home-built multimodal scanning microscope, based on off-the-shelf components and allowing maximum accessibility to the beam paths. We demonstrate CARS/SRS imaging of polymer beads and of leaves of the *Elodea* aquatic plant.

## 2 Material and methods

Figure 1(a) shows the architecture of the compact multi-branch fiber-format laser source that generates the multi-colour pulses required for the different NLO microscopy modalities. It is based on a mode-locked Erbium:fiber oscillator working at 40-MHz repetition rate, which feeds three independent Er-doped fiber amplifiers (EDFAs), each producing 350-mW average power at 1560-nm central wavelength. In this way, all three EDFA outputs are inherently synchronized and phase coherent [30]. Further, they are compressed to nearly transform-limited sub-100-fs duration by individual pairs of silicon prisms. Two of the EDFA outputs (referred to as 'Arm 1' and 'Arm 2' in Fig. 1(a)) are coupled to highly nonlinear fibers (HNLFs) that considerably broaden the laser spectrum, generating side-lobes at both longer and shorter wavelengths with respect to the input fundamental one. The longer wavelength peak is a soliton, produced mainly through self-frequency shift due to intra-pulse Raman scattering. On the other hand, the shorter wavelength peak is propagating in the normal dispersion regime, also phase-matched to the soliton. The spectral positions of the two peaks sensitively depend on the overall dispersion of the pulse coupled to the HNLF. Acting on the silicon prism-pair, the position of soliton peak can be continuously tuned and precisely controlled between 1700 and 2100nm.

The third EDFA output [referred to as 'Arm 0' in Fig. 1(a)] at 1560 nm is sent to a non-linear crystal to generate the pump pulses for the CRS processes at a fixed wavelength of 778 nm. We employed a 10-mm-long MgO-doped periodically poled lithium niobate (PPLN) crystal with a poling period of 19.3 μm. It generates a narrow-bandwidth (15-cm$^{-1}$ linewidth) second harmonic, according to the spectral compression technique [31], with 120-mW average power [see green spectrum in Fig. 1(b)]. In the same manner, the redshifted soliton output of one of the HNLFs (in 'Arm 1') is frequency doubled in a 10-mm-long PPLN crystal having a fan-out grating design, spanning the poling period range of 26–33 μm. It generates tunable Stokes pulses for the CRS modalities from 950 to 1050 nm [see red spectra in Fig. 1(b)] with power up to 10 mW and linewidth ranging between 18 and 30 cm$^{-1}$. Tuning of the Stokes wavelength is achieved simply by transversely translating the fan-out crystal using a motorized stage, calibrated to guarantee a rapid (within a fraction of a second) and reproducible selection of the vibrational Raman shift. The pump-Stokes frequency detuning in the range 2330-3330 cm$^{-1}$ fully covers the CH-vibrational region, which is the most commonly used in CRS microscopy. Before entering the microscope, the pump and Stokes pulses are temporally matched by a mechanical delay line and then collinearly combined by a dichroic beam splitter (Semrock, LP02-785RS-25). For the SRS modality, the pump beam is modulated at 1 MHz by an acousto-optic modulator placed in the beam before the combiner. The narrowband pump or Stokes pulses can individually be utilized as an excitation pulse for TPEF or SHG/THG experiments as per demand of the sample. Furthermore, a part of the compressed sub-100 fs output of the EDFA at 1560 nm just before the PPLN (in Arm 0) can be split and used as an excitation pulse for THG microscopy.

The third branch [referred to as 'Arm 2' in Fig. 1(b)] is used, by properly adjusting the silicon prism pair before the HNLF, to generate a broadband spectrum in the 840- to 1100-nm wavelength region. After compression in a SF-10 prism pair, we obtain sub-20-fs pulses, which can be used either as Stokes pulses to implement broadband CARS/SRS configurations [32, 33] or as excitation pulses for TPEF or SHG experiments performed on the microscope. In the following, we will present experimental results obtained coupling only 'Arm 0' and 'Arm 1' beams to the multimodal NLO microscope.

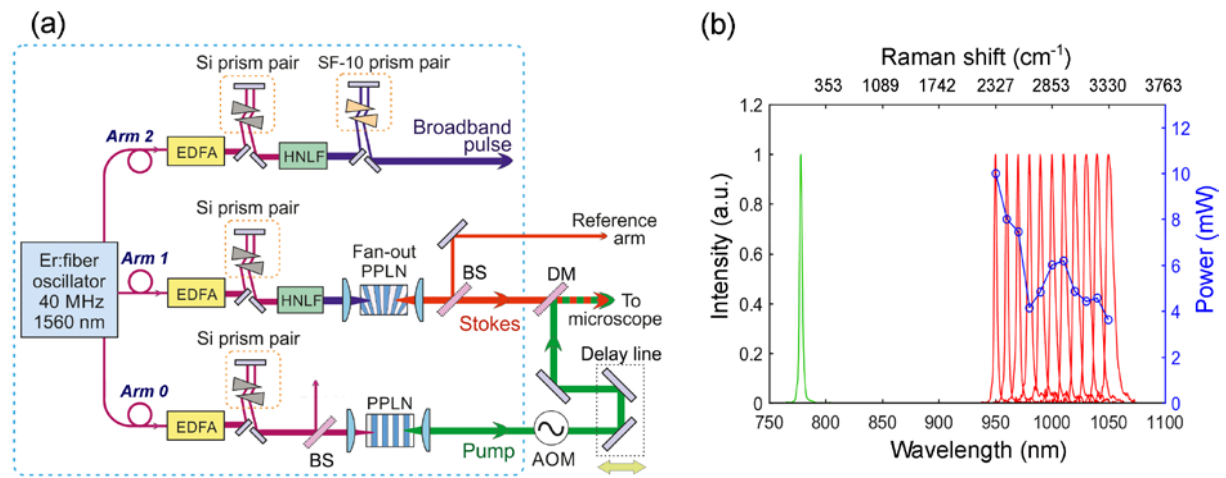

*Figure 1: (a) Schematic of a multi branched Er: fiber laser source for multimodal nonlinear microscopy. AOM, acousto-optic modulator; DM, dichroic mirror; PPLN, periodically poled lithium niobate; EDFA, erbium-doped fiber amplifier; HNLF, highly nonlinear fiber; BS, beam splitter; Si, silicon. (b) Spectra of narrowband pump (Arm 0) and tunable narrowband Stokes (Arm 1) beams; The curve in blue indicates the wavelength dependent average power of the Stokes beam arriving on the sample.*

We have developed a simple and low-cost multimodal NLO laser-scanning microscope, following the design in Fig. 2. A pair of galvanometric mirrors (Cambridge Technology: ProSeries II Scan Head-10mm) raster scans the input beams over the sample area at the focus of the microscope objective. For this purpose, two telecentric lenses, the scan and tube lenses (Thorlabs: LSM05 BB, EFL 110 mm), are arranged in a standard 4-f configuration to form an image relay system between the scan head and the back focal plane of the focusing objective. This objective (Zeiss 100×, NA = 0.75, WD= 4 mm, field number 25) provides a flat field-of-view (FOV) of 250×250 µm$^2$ area and a lateral spatial resolution of approximately 1 µm. The samples are mounted on a computer-controlled X-Y-Z stage, composed of a motorized X-Y-scanning stage (Standa: model 8MTF-102LS05) and a motorized Z-translation stage (Mad City Labs Inc.: model MMP1). This stage enables us to bring the desired sample area under the FOV of the focusing objective and to acquire images of large sample areas by tiling consecutive measurements. Alternatively, it can also be utilized as a raster scanning stage for acquiring images of the sample without scanning the beams. This method is particularly important for microscopic techniques that are sensitive either to polarization, such as balanced-detection Raman induced Kerr-effect microscopy [34], or to beam pointing, such as broadband Fourier-transform SRS using a birefringent interferometer [33] or broadband SRS spectroscopy using a photonic time stretcher [35].

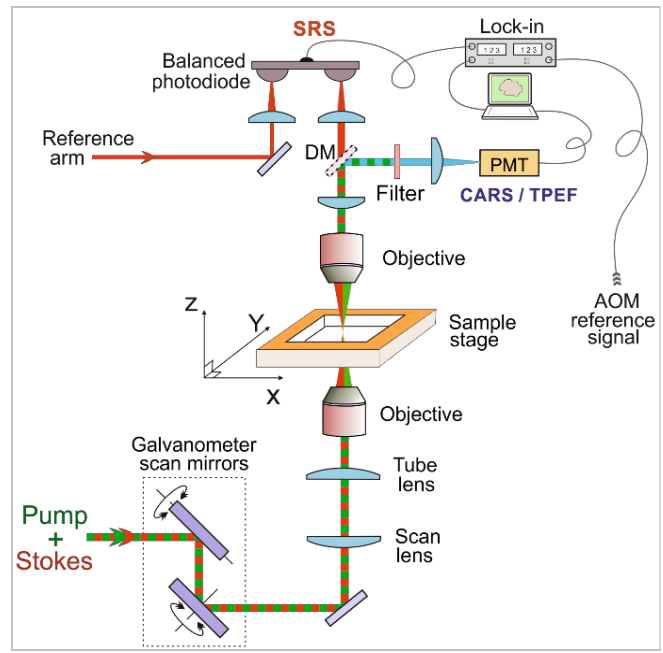

*Figure 2: Sketch of the multimodal NLO laser-scanning microscope, depicting detection schemes for SRS and CARS/TPEF modalities in forward direction; DM, dichroic mirror; PMT, photomultiplier tube; AOM, acousto-optic modulator.*

After the sample, the generated forward CARS, SRS and TPEF signals are collected by a second identical microscope objective. A dichroic beam splitter transmits the Stokes beam, bearing the SRG signal, and reflects the shorter wavelengths, containing CARS/TPEF signals and the pump beam itself. A photomultiplier tube (Hamamatsu: R3896) reads CARS or TPEF signals after spectral filtering by applying appropriate bandpass interferential filters. A balanced photodiode (Thorlabs: PDB210A/M) and a high-speed lock-in amplifier (Zurich Instruments: HF2LI) detect the transmitted SRG signal by a modulation-transfer scheme. A reference beam for the balanced photodiode is obtained by splitting a small portion of the Stokes beam just before the combiner (shown in Fig. 1(a) as 'Reference arm'). A set of lenses in 4-f configuration are placed after the sample to conjugate the back aperture of the collimating objective with the detectors, thus guaranteeing a proper collection of light within the active areas of the detectors. Our NLO microscope can also be readily reconfigured to work in *epi* (back-scattering) detection, just by placing a dichroic beam splitter between the tube lens and the focusing objective to reflect epi- CARS, TPEF, SHG or THG signals, to be collected by their respective PMTs/detectors.

## 3 Results and discussion

We first tested our NLO microscope on a sample made of a mixture of two different polymer beads, namely poly-methyl methacrylate (PMMA) with 6−µm diameter and polystyrene (PS) with 3−µm diameter, dispersed on a glass substrate. We simultaneously acquire CARS and SRS images (with 200×200 pixels) in galvanometric-mirrors scanning mode of an 80×80 µm$^2$ area of the sample with 3-ms pixel dwell time. The average pump and Stokes powers on the sample are 12 mW (modulated at 1 MHz) and 5 mW, respectively. Figures 3(a) and 3(b) report the CARS and SRS images at pump-Stokes frequency detuning of 2953 cm$^{-1}$, in resonance with PMMA vibrational response. Figures 3(c) and 3(d) show the CARS and SRS images, at pump-Stokes frequency detuning of 3060 cm$^{-1}$, mainly in resonance with PS response. Figures 3(e) and 3(f) are the overlay images of CARS signals (a, c) and of SRG signals (b, d), respectively. One can clearly see in false colour mapping (PMMA in red and PS in green) the chemical selectivity of CRS microscopy. We have also acquired the CARS/SRS spectra of PMMA and PS [Fig. 3(g) and 3(h)] by parking the microscope at points 'A' and 'B' (as shown in the panels a-d), respectively. They were recorded by scanning the Stokes wavelength with the help of the motorized translation stage of the fan-out PPLN in 'Arm 1'. SRS spectra [Fig. 3(h)], being inherently free from NRB, display the main PMMA resonance at 2953 cm$^{-1}$ and the two peaks of PS at 3060 cm$^{-1}$ and 2910 cm$^{-1}$ with symmetrical lineshape and with central positions in good agreement with literature. As expected, CARS spectra [Fig. 3(g)], suffering from interference with NRB, show characteristic dispersive lineshapes and blue-shifted central positions of the peaks.

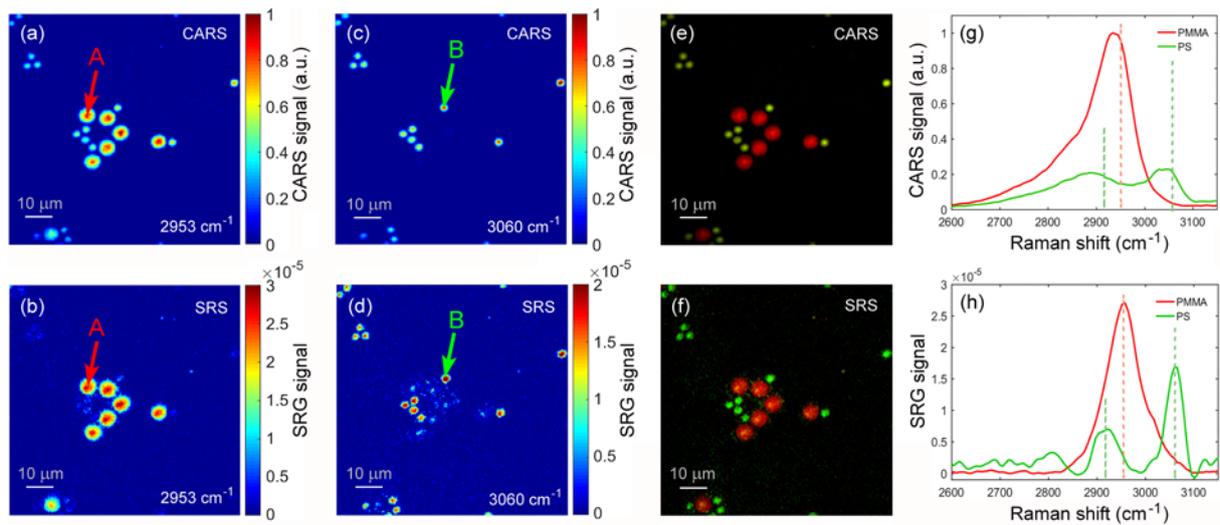

*Figure 3: Images (200×200 pixels) of a mixture of 6−μm PMMA beads and 3−μm PS beads for CARS (a,c) and SRS (b,d) modalities, at 2953 cm$^{-1}$ (a,b) and 3060 cm$^{-1}$ (c,d) with 3-ms pixel dwell time. (e,f): overlay images of CARS (a,c) and SRS (b,d) images. CARS (g) and SRS (h) spectra taken at points 'A' (in red) and 'B' (in green).*

Being NLO techniques, both CARS and SRS have three-dimensional (3D) sectioning capabilities. We performed 3D-CARS chemical imaging for our composite sample of polymer beads. To this purpose, we acquired a set of CARS images (over 50×50 μm$^2$ area with 125x125 pixels) at different z positions within the sample at two frequency detunings: 2953 cm$^{-1}$, in resonance with PMMA, and 3060 cm$^{-1}$, in resonance with PS. In particular, we varied the axial (z) position of the sample throughout the laser focus over 30-μm travel range with a step size of 0.5 μm, for a total of 61 steps for each of the two vibrational frequencies. Figure 4(a) shows the overlay of two CARS images collected at a particular z position for the two frequency detunings (red, PMMA and green, PS). Figure 4(b) shows the resulting 3D reconstruction of the sample volume, employing the entire acquired dataset.

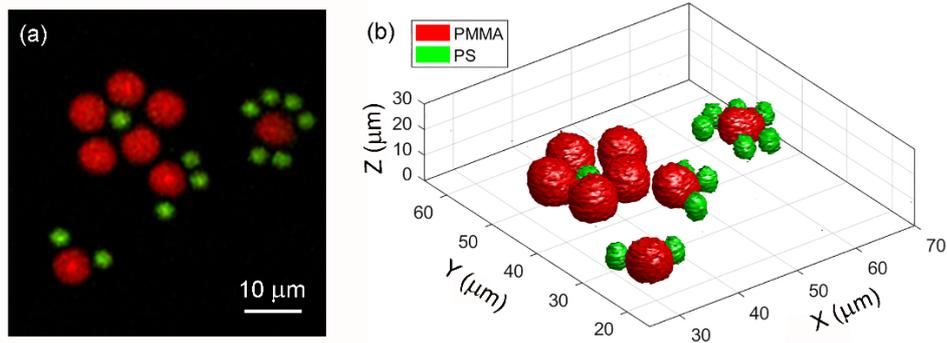

*Figure 4: (a) Overlay of two CARS images collected at a fixed z position (acquired at 2953 cm$^{-1}$ in resonance with PMMA, red, and 3060 cm$^{-1}$, in resonance with PS, green) with 200 µs pixel dwell time. (b) 3D reconstruction of the sample volume by rendering the two sets of 61 images for the two resonances.*

We further acquired CARS images of the same polymer beads sample at several pump-Stokes frequency detunings in the 2600 to 3200 cm$^{-1}$ range, at 10 cm$^{-1}$ intervals. We mapped a sample area of 50×50 µm$^2$ with 125×125 pixels, at a fixed z axial position. We then performed multivariate curve resolution (MCR) analysis [36, 37] on the acquired 125×125×61 data matrix, obtaining two dominant spectral components related to PMMA and PS [shown in Fig. 5(c)]. The concentration maps associated with these two components are plotted in Fig. 5(a) and 5(b), respectively.

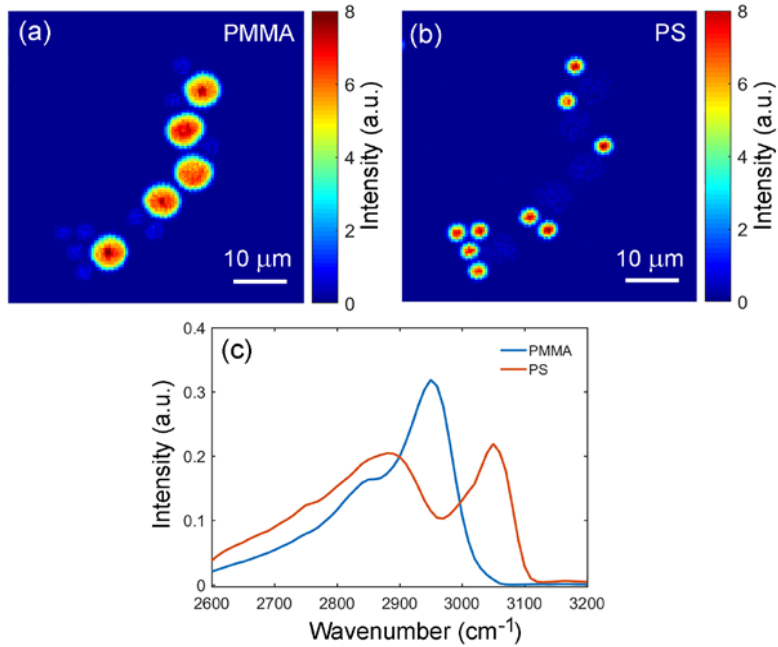

*Figure 5: MCR analysis of the recorded hyperspectral CARS dataset (acquired with 200-µs pixel dwell time) of the polymer beads sample. (a,b) Concentration maps of the first and second components, associated to PMMA and PS, respectively. (c) Retrieved spectral shapes of the two components (PMMA, blue, and PS, red).*

To demonstrate the multimodal imaging capability of the microscope on a biological sample, we imaged an unprocessed fresh leaf of *Elodea* aquatic plant. A small part of the leaf with a drop of water is sandwiched between two thin microscope glass slides. Figure 6(a) shows an SRS image (200×200 pixels, 100×100 µm$^2$ area) of the sample obtained at a Raman shift of 2890 cm$^{-1}$ (pump, 778 nm and Stokes 1004 nm), mainly targeting the cellulose, a constituent molecule for plant cell walls. The (thicker) cell walls mostly made of cellulose can clearly be seen. We also performed TPEF microscopy of the same sample area targeting mainly the intracellular Chlorophyll *a*, excited by two-photon absorption of pump pulses at 778 nm. This wavelength is well suited for the purpose, because Chlorophylls have negligible linear absorption at the 778-nm pump wavelength [38,39]. Even if the Stokes wavelength at 1004 nm does not contribute to the TPEF process, in combination with the pump it can generate CARS or four-wave-mixing signals that can spectrally overlap with TPEF. To avoid these processes, we merely go out of time overlap between the pump and Stokes pulses. Under this situation, the TPEF image acquired in forward direction on the PMT in the spectral window 600-700nm is shown in Fig. 6(b). Round shaped Chlorophyll-rich areas can be seen within the intracellular volume. Figure 6(c) is the overlay of the two, SRS (in red) and TPEF (in green) images.

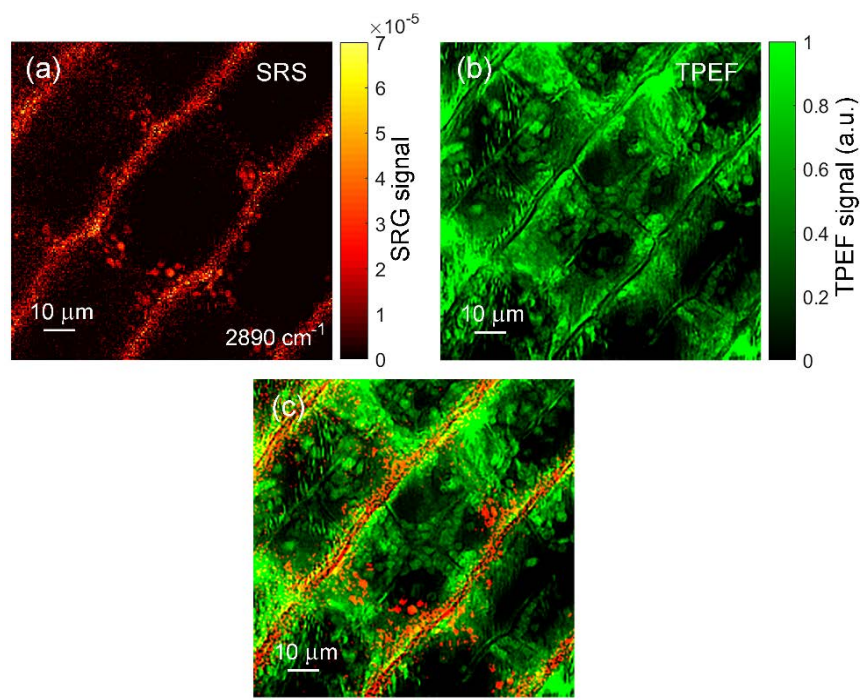

*Figure 6: 200×200 pixels images of a 100×100 μm² area of an unprocessed fresh leaf of 'Elodea' aquatic plant; (a) SRS image acquired at pump-Stokes frequency detuning of 2890 cm$^{-1}$, mainly targeting cellulose, a major constituent of plant cell walls;(b) TPEF image acquired with pump pulse excitation, depicting round shaped intracellular chloroplasts, (c) An overlay of the two images.*

## 4 Conclusions

We have presented a multimodal NLO laser-scanning microscope, integrating CARS, SRS and TPEF imaging modalities on a single platform. The system combines a compact, turnkey and potentially low-cost fiber-format excitation laser with a microscope, developed in house using off-the-shelf components, which offers full accessibility to the optical path and ensures its easy re-configurability and flexibility. This system, while performing very closely to state-of-the-art setups, provides a cost-effective alternative to the commercially available expensive multimodal nonlinear optical imaging systems. Experimental results presented here on CARS/SRS and TPEF microscopy in forward direction of polymer and plant samples, are quite promising in view of applications of this microscope to mainstream biomedical problems.

**Conflict of Interest**

The authors declare no competing financial interest.

**Acknowledgements**

This work is supported by European Research Council: Consolidator Grant VIBRA (ERC-2014-CoG No. 648615), and Advanced Grant STRATUS (ERC-2011-AdG No. 291198).